\documentstyle[times,epsfig]{jaa}
\newcommand{\lboo}{$\lambda$~Bootis}
\newcommand{\dscuti}{$\delta$~Scuti}
\begin{document}
\title[Seismology of non-magnetic CP stars]{Using the seismology
of non-magnetic chemically peculiar stars as a probe of dynamical 
processes in stellar interiors}
\author[S. Turcotte]%
       {Sylvain Turcotte\thanks{e-mail:turcotte@apollo.ubishops.ca} \\ 
        Bishop's University, Lennoxville, Canada, J1M 1Z7}
\setcounter{figure}{0}

\maketitle
\label{firstpage}

\begin{abstract}
Chemical composition is a good tracer of hydrodynamical processes that 
occur in stars as they often lead to mixing and particle transport.
By comparing abundances predicted by models and those observed in stars
we can infer some constraints on those mixing processes. 
As pulsations in stars are often very sensitive to chemical composition, we can 
use asteroseismology to probe the internal chemical composition of stars where no
direct observations are possible.
In this paper I focus on main sequence stars Am, \lboo, and HgMn stars and discuss
what we can learn of mixing processes in those stars from seismology.
\end{abstract}


\section{Introduction}
\label{sec:intro}
Chemically peculiar stars (CP stars) are stars in which at least one but typically
several elements are significantly over- or under-abundant with respect
to what is considered a normal composition for such stars. In population~I
main-sequence stars this standard is the Sun. 

Chemically peculiar stars come in many forms throughout the HR diagram, and 
the physical mechanism(s) responsible for those anomalies vary in different regions 
of the HR diagram. In giant stars for example episodes of deep convection 
mix matter that has undergone nucleosynthesis with matter from cooler regions 
unaffected by nuclear processes (see Busso et al. 1999 for example). In cool population~I and
population~II stars, turbulent mixing may lead to the destruction 
of light elements such as lithium from the surface layers of the stars (Michaud \& Charbonneau~1991).
The Sun for example is depleted of lithium by more than two orders of magnitude. 
Accretion of material onto the surface of stars or mass loss by a stellar wind have also 
been shown as means to produce chemically anomalous stars 
(Venn \& Lambert~1990; Proffitt \& Michaud~1989).  
Finally, microscopic diffusion 
due to gravity and radiation pressure acting differently on 
each atomic species is another way to build abundance anomalies in some types of
stars (see a recent review by Vauclair~(2003) and references therein). 
In the following we will focus on diffusion and accretion as they are the
mechanisms that are dominant in main sequence CP stars of low to intermediate mass.

The chemical anomalies observed at the surface of CP stars extend in their interior.
The abundance profiles depend on the interplay of diffusion, of other processes
promoting stratification such as accretion and of mixing processes. It is these
abundance profiles that we may be able to probe by studying the pulsations of those
stars. If the basic physical processes leading to the formation of inhomogeneities are
modeled accurately, we can then use the information on internal composition provided
by pulsations to put constraints on the mixing processes that oppose the changes
in composition.

In what follows we will see how chemical anomalies and pulsations are connected
more specifically in main-sequence A and B-type stars where we find a large 
variety of both CP and variable stars.  We will consider only non-magnetic stars 
to avoid the complications posed by the largely unknown strength and
geometry of the magnetic fields in stellar interiors. Constraints on mixing processes can
only be inferred when the other processes leading to the evolution of the composition
can be modeled with no free parameters.
We will concentrate on Am and HgMn stars, main sequence A and B type stars respectively, in
which diffusion is the governing process and on \lboo\ stars that are thought
to be formed through accretion of metal depleted circumstellar matter.

\section{Abundance stratification in main-sequence stars}
\label{sec:process}

The governing equation for the evolution of abundances in a star is 
\begin{equation}
  {\partial X\over \partial t}  = \Bigl[ (D_i+D_T) {\partial X\over \partial M} 
    + (v_i+v_w+v_{\rm hydro}) X \Bigr] - \lambda_{\rm sink} X + \lambda_{\rm source} \, ,
\end{equation}
where $X$ is the mass fraction of a species. $D_i$ and $v_i$ are the diffusion coefficient
and velocity for that species, $D_T$ is the turbulent diffusion coefficient, $v_w$ is
a velocity due to mass loss (if positive) or accretion (if negative), $v_{\rm hydro}$ is
a velocity due to large scale motions of matter (meridional circulation f. ex.),
$\lambda_{\rm sink}$ is a sink term (non zero at the surface in the case of mass loss),
and $\lambda_{\rm source}$ is a source term (non zero at the surface in the case of accretion).

The important processes that lead to abundance stratification (and surface 
abundance anomalies) and the competing mixing mechanisms contribute differently
in each star depending on its temperature, age, mass, and rate of rotation. 
Accretion on the other hand, when present, is an external process practically 
independent of the properties of the star. 

\subsection{Mixing}
\label{sec:mixing}

Several processes lead to mixing in stars, in some cases through large
scale motion that transport matter throughout the interior, in other cases through
local motions such as turbulence. The most common and most important of
the mixing mechanisms is convection.
In so-called {\sl standard} stellar models, convection is typically
the only mixing mechanism considered.
Amongst the other mixing processes the most important are 
convection overshoot and rotational induced mixing.
Rotation leads to large scale motions meridional circulation but also to turbulent
motions induced by rotational shear or as the result of the transfer of angular momentum
from the core toward the surface. In stellar models turbulent mixing due to convection or
other forms of turbulent motions is accounted for through the 
turbulent diffusion coefficient ($D_T$ in Eq.~1). 
In the models discussed below, $D_T$ for convection is arbitrary but must be 
very large, while the functional form for the combined effect of all other mixing
mechanisms is an ad hoc power law (Richer et al. 2000).  

Convection typically occurs in regions of partial ionization of HI and HeI at around 10~000~K
and of HeII at around 40~000~K.
In non-magnetic stars with surface temperature of less than 10~000~K, the outer regions 
of the star, including the photosphere, are convective. The HI+HeI and HeII convection zones
are expected to be linked by convection overshoot. As a result the whole outer envelope
of these stars are completely mixed. Mixing is also expected to extend below the HeII 
convection zone because of overshoot or turbulence. The region in which the star is fully
mixed extends to a certain depth. This mixed region is named the Superficial Mixed Zone (SMZ)
in the remainder of this paper. The extent of the SMZ may be determined by seismology.

In cool stars, the photospheric abundance measured spectroscopically are identical
to the abundances at the deepest point of the SMZ. Surface abundances are determined by 
particle processes that occur relatively deep in the interior.
In stars hotter than 10~000~K, the outermost regions are not convective and therefore 
may not be mixed. Stratification can therefore occur in the photosphere. 
As a result, the photospheric abundances can be completely different from
the abundances in the star's interior.

\subsection{Diffusion}
\label{sec:diffusion}

Chemical species drift relative to each other as the net forces acting on each species
vary depending on their atomic properties. Essentially, it is the competition between gravity 
and radiation pressure that determines if an element will levitate toward the surface or 
sink toward the center of the star. While gravity is essentially constant in the outer parts of 
stars, radiation pressure goes through peaks and troughs depending on the local temperature 
and density. Where radiation pressure is larger than gravity, an element will drift outward,
but will sink if gravity is larger. Fig.~\ref{fig:mixing} shows a cartoon that illustrates
the effect of diffusion on the chemical profiles and the effect of the depth of the SMZ
on surface abundances. In the middle panel, the base of the SMZ occurs where gravity is
larger than radiation pressure so the abundances in the SMZ are lower than initially.
In the bottom panel, the deeper SMZ yields an increase in the surface composition but a 
smaller anomaly due to the larger mass of the SMZ.

The effect of diffusion on abundances can be calculated rather accurately from first 
principles. As a result, stars in which diffusion is the dominant process in creating
abundance gradients are very useful tools to study mixing processes.

\setcounter{figure}{0}
\begin{center}
\begin{figure}
 \epsfig{file=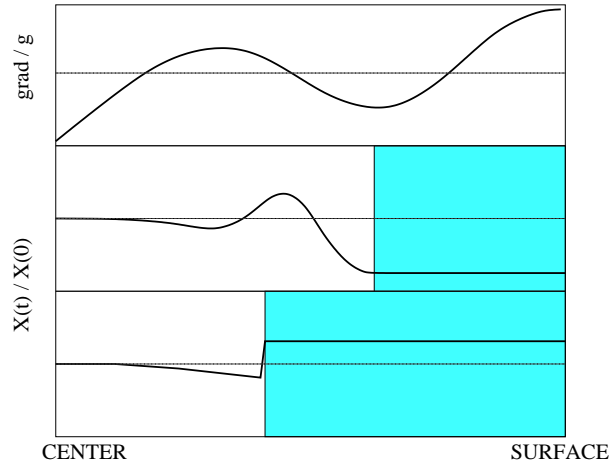,width=8cm}
 \caption[]{The top panel is a cartoon representation of the ratio between the 
acceleration due to radiative pressure ($g_{\rm rad}$) and the gravitational acceleration ($g$), where 
the horizontal dotted line shows a ratio of one. The middle panel shows the evolution of
the abundance profile as the result of the $g_{\rm rad}/g$ profile shown above. The shaded 
region on the right is the Superficial Mixed Zone (SMZ) in which the abundance is homogenized
by mixing. The bottom panel shows the evolution of the same element but for a deeper SMZ. 
\label{fig:mixing}}
\end{figure}
\end{center}

\subsection{Accretion}
\label{sec:accretion}

Accretion does not depend directly on intrinsic properties of the accreting star
but the material falling on the surface of the star must be mixed (diluted) in the
entire SMZ. The composition in the mixed zone evolves on a time scale
defined by the ratio of the mass of the mixed zone to the accretion rate 
($\tau=M_{\rm SMZ}/\dot{M}$; neglecting here the effect of other processes at 
the base of the mixed zone).  The depth in the star that the accreted material will
reach depends on the depth of the SMZ, but also on the advection of the accreted 
matter in the star at a velocity of $v=\dot{M}/4\pi\rho r^2$ (Charbonneau 1991).

Given a large enough rate of accretion (i.e. a short $\tau$ compared to the evolution timescale), 
the abundances in the SMZ will be identical 
to those of the accreted matter. Once the accretion episode ends the signature of accretion
rapidly disappears (see detailed models in Turcotte \& Charbonneau~1993).
The abundance profiles will depend on the SMZ, on $\dot{M}$, and on the cumulative
mass of the accreted material. None of which can be measured directly.
As a result it may be difficult to find the depth of the SMZ uniquely for a given star.

\section{Composition anomalies and their effect on pulsations}
\label{sec:opacity}

Different elements dominate the opacity at different depths in a star. Hydrogen and
helium contribute most of the opacity at low temperature, iron dominates
in hotter regions. In CP stars, the opacity profile will be different than in
chemically homogeneous stars. The opacity might be smaller if the elements that dominate
at a given temperature become less abundant, and inversely, might increase 
elsewhere as some elements accumulate where they play an important role. 

In Fig.~\ref{fig:opacity}, the contribution of several elements to the opacity is shown in the
case of a chemically homogeneous star of solar composition and in a star in which the abundances
have been changed by diffusion. One can notice that the contributions of helium and of
CNO have been reduced in the outer regions of the star while the contributions of
iron-peak elements have increased. As a result, the driving due to helium is smaller
in stars with diffusion than in normal stars, while the driving due to heavy elements
is enhanced. 
\setcounter{figure}{1}
\begin{figure}
 \epsfig{file=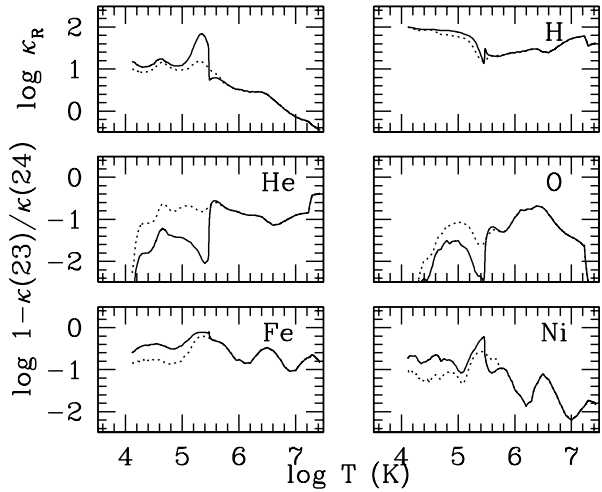,width=8cm}
 \caption[]{The relative contribution of individual elements to the Rosseland mean opacity 
is shown for 5 elements when the chemical composition is uniformly solar (dotted line) and
when diffusion as occurred (solid line). The top left panel shows the logarithm of the
Rosseland mean opacity in both cases. 
In the five other panels the logarithm of the ratio of the opacity including 
all elements in the star ($\kappa(24)$) and the opacity calculated without the contribution of
the element labeled ($\kappa(23)$) is plotted. If an element provides all the opacity, then the
curve for that element would reach one. The model shown is a 4 M$_\odot$ star model of 100~Myr 
(see Turcotte \& Richard 2005, submitted).  \label{fig:opacity}}
\end{figure}

In stars in which accretion is ongoing, the effect on driving will be determined by
the content of the accreted material and the depth in the star reached by the 
accreted material. In \lboo\ stars, the driving due to helium will be unchanged as
helium is expected to be present in the accreted material and the helium driving
region is inside the surface convection zone. The driving due to heavy elements will
be changed only in certain circumstances.

\section{Inference of mixing processes in CP stars}
\label{sec:infer}

The ideas above have been applied to three types of chemically peculiar stars
of the main sequence, the A-type Am and \lboo\ stars, and the B-type HgMn stars. 
Although similar in many ways these three types of stars
are nonetheless different in important ways: When variable, Am and \lboo\ stars are delta Scuti type 
pulsators (p-modes driven by helium) while pulsating late-type B stars are 
g-modes excited by the opacity of heavy elements;
Am and HgMn stars are slowly rotating while \lboo\ stars are moderately rapid rotators;
and, finally, because of differences in the position of convection zones, the 
photospheric composition of Am and \lboo\ stars is the same as that of the driving region
while it's not the case for HgMn stars because of diffusion in their atmosphere.
The expected differences between these three types of stars are illustrated in 
Fig.~\ref{fig:stars}. 
\setcounter{figure}{2}
\begin{figure}
 \epsfig{file=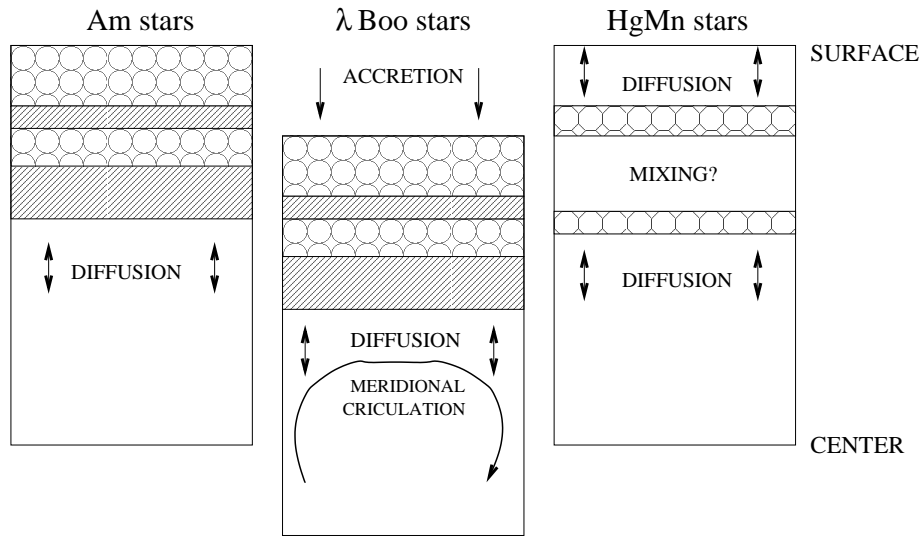,width=4.8in}
 \caption[]{The cartoons (not to scale) illustrate the expected processes important in the evolution 
of the abundances of the three types of stars discussed in the paper.
The regions filled with circles represents convection zones, the regions filled by oblique
lines are non convective regions but otherwise mixed by other turbulent processes.
The presence or absence of non-convective mixed regions in HgMn stars is not known. 
The deeper convection zone in HgMn stars appears only when iron-peak elements have accumulated 
in that region as the result of diffusion.
\label{fig:stars}}
\end{figure}

\subsection{Application to Am stars}
\label{sec:Am}

Models of non-rotating A stars were computed to test whether variable Am stars
can be driven by the opacity of helium as in 
\dscuti\ stars (Turcotte et al. 2000). As suggested in Fig~\ref{fig:opacity},
as the helium settles out of the driving region as a result of diffusion
the driving in Am stars is much lower than in chemically normal stars.
As the helium driving region is inside the superficial convection, and thus inside the SMZ,
the level of depletion of helium in the driving region is a direct function 
of the depth at which the SMZ extends, the deeper is the SMZ the higher the 
driving will be.

The blue (hot) edge of the instability zone for Am stars is in direct relation
with the abundance of helium in the driving region as is shown in Fig.~\ref{fig:Am}. 
Therefore, we can infer the depth of mixing from the instability
region for Am stars.
The boundary of the zone has not been explored in detail yet, partly because of 
the computational cost to do so. An additional difficulty is that the initial
helium content of individual Am stars need not be identical and so the 
depth of mixing may not be a unique solution for all variable Am stars.
Nevertheless, the depth of mixing found through seismology is consistent with what was found 
from modeling the surface composition (Richer et al. 2000).

\setcounter{figure}{3}
\begin{figure}
 \epsfig{file=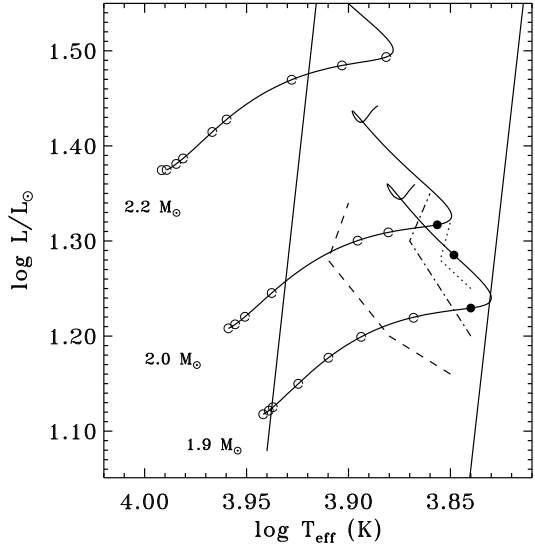,width=8cm}
 \caption[]{The HR diagram shows the approximate boundaries of the classical 
instability strip (oblique lines), the evolutionary paths of 1.9, 2, and 2.2 M$_\odot$ 
stars undergoing diffusion. The filled circles represent models where at least one 
unstable pulsation mode was found, the open circles show models that are stable.
The three other lines show the approximate boundaries of the instability region
for variable Am stars when the SMZ is shallow (dotted line), best fit (dash-dotted),
or deep (dashed). \label{fig:Am}}
\end{figure}

\subsection{Mixing and accretion in $\lambda$~Bootis stars}
\label{sec:lboo}

In the accretion model the light elements (H, He, CNO) in the SMZ are replenished 
by the infalling gas while most other elements are depleted by a factor of 10
in the accreting matter (Turcotte \& Charbonneau~1993). 
As a result, one would not expect to see any measurable effect
on the pulsations of \lboo\ stars in comparison to \dscuti\ stars if the SMZ is not 
significantly deeper than the surface convection zone of normal A stars.
If, however, the SMZ extends significantly deeper, to the depth at which the opacity
of heavy elements drive pulsations, an observable signature might be generated.
Fig.~\ref{fig:lboo} shows the relative change in frequency for radial modes ($\ell=0$)
of radial order ($n$) of 1 to 14 with respect to a chemically homogeneous model.
We see that the differences are significant only if the chemical anomalies extend to a
depth where the low metallicity of the accreted matter can make a large difference in
the opacity ($\log T>5.2$). Models of Am stars siggest this should be the case, and as
\lboo\ stars are faster rotators one may expect more mixing in \lboo\ stars than in Am stars. 

Unfortunately, using frequencies to learn about the structure of \dscuti\ type stars
as yet to be fruitful due to problems in mode identification and 
in dealing with rotation theoretically.
Consequently, relying on the small frequency differences between the different models
to infer the depth of the SMZ promises to be challenging.

\setcounter{figure}{4}
\begin{figure}
 \epsfig{file=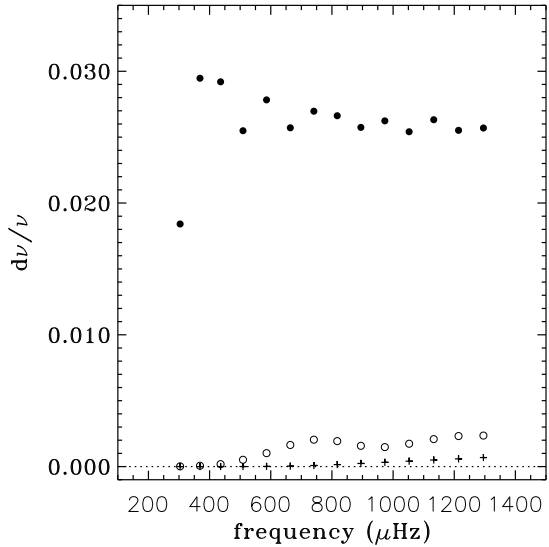,width=8cm}
 \caption[]{The relative difference in frequencies of radial modes of order 1 to
14 for models that are metal poor (except for CNO and S) to a depth where the tenperature is
$\log T=4.4$ (crosses), $\log T=5.0$ (open circles), and $\log T=5.8$ (filled circles) 
with respect of  a chemically homogenous model
are shown at the frequency of the reference model. 
The SMZ in the last model is $10^6$ more massive than in the first. \label{fig:lboo}}
\end{figure}

\subsection{What is the matter with HgMn stars?}
\label{sec:HgMn}

In pulsating B stars the driving is due to iron-peak elements (Pamyatnykh~1999).
As HgMn stars are slowly rotating, the current theoretical expectation is that diffusion 
will occur in the interior as it does in the atmosphere 
(as the large abundance anomalies measured there attest). 
Models show that diffusion leads to an accumulation of iron-peak elements 
in the driving region. The increase in abundance of those elements naturally leads to
an increase in opacity, which in turn leads to an enhanced driving.
Fig.~\ref{fig:HgMn} shows the logarithmic opacity gradients for models of 
a 4~M$_\odot$ where the depth of the SMZ is varied. The increase of the peak at 
$\log T\approx 5.2$ leads to an increase in the derivative of these opacity gradients 
which means that the driving does increase (see Shibahashi in these proceedings).

However, HgMn stars are not known to be variable whereas chemically normal stars
of the same mass and evolutionary state are variable (those are the Slowly Pulsating B stars).
The disagreement between the models and observations suggest that the models are deficient
in some ways. A solution may come from introducing new physics in the models such as 
selective mass loss, but another possibility may simply be that the models are not accurate enough
in the superficial regions. Indeed, the models discussed here assume for reasons of
numerical stability that the stars are fully mixed from 
the photosphere to a depth past the mode driving region which is not realistic 
(Turcotte \& Richard 2005, submitted). Only through
further modeling combined with seismic data will we establish whether HgMn stars
are really stable or variable at very low amplitude and what implication, if any, this has
on dynamical processes occuring in these stars.

\setcounter{figure}{5}
\begin{figure}
 \epsfig{file=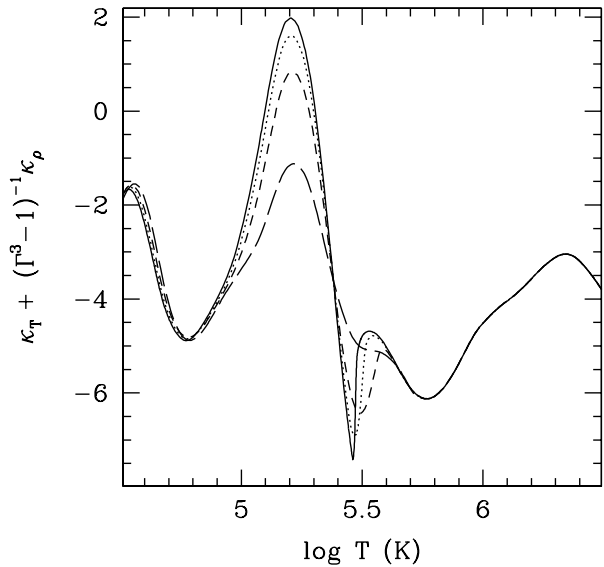,width=8cm}
 \caption[]{The sum of the logarithmic derivatives of opacity ($\kappa_T=d\log\kappa/d\log T$,
$\kappa_\rho=d\log\kappa/d\log\rho$) is shown for
four 4~M$_\odot$ models with more or less mixing versus the temperature in the star. 
The peak at around $\log T = 5.2$ is due to the opacity of iron-peak elements. It is 
larger in models where mixing is lower and abundance anomalies are allowed to be larger.
The pulsations are driven in regions where 
$d/dr[ \kappa_T + \kappa_\rho/(\Gamma_3-1) ] > 0$.
\label{fig:HgMn}}
\end{figure}

\section{Conclusions}
\label{sec:conclusion}

Using variability to infer mixing is especially useful in stars for which 
surface abundances is not a good indicator of internal composition, such
as HgMn stars, and, to a certain degree, \lboo\ stars.
In Am stars on the other hand, variability is an indicator of the abundance 
of helium, which is useful as it cannot be seen in spectra of these stars.

Much more work is required to refine this tool. At this time 
only the depth of the superficial region where abundances are homogeneous (the SMZ)
can be constrained in Am stars. The exact processes that lead to that mixing 
cannot be isolated nor can the exact profile of the turbulent diffusion coefficient
be determined.  Nevertheless, this tool remains our best hope to study the
internal chemical composition of stars of the upper main-sequence as models improve and as the
data from MOST, CoRot and other missions hopefully provide us with low amplitude
modes in Am, \lboo, and HgMn stars.

\end{document}